\documentclass[floats,aps,twocolumn,showpacs]{revtex4}
\usepackage{graphicx}
\usepackage{color}
\usepackage{amsmath}
\begin{document}
\newcommand{\BSCCO}{{Bi$_2$Sr$_2$CaCu$_2$O$_{8+x}$}}
\newcommand{\YBCO}{{YBa$_2$Cu$_3$O$_{7-\delta}$}}
\def\k{{\bf k}}
\def\q{{\bf q}}

\title{Fourier transform spectroscopy  of $d$-wave quasiparticles  in
the presence of atomic scale pairing disorder}
\author{Tamara S. Nunner, Wei Chen, Brian M. Andersen, Ashot
Melikyan, and  P. J. Hirschfeld}

\affiliation{Department of Physics, University of Florida,
Gainesville, FL 32611, USA }
\date{\today}

\begin{abstract}
The local density of states power spectrum of optimally doped
Bi$_2$Sr$_2$CaCu$_2$O$_{8+x}$ (BSCCO) has been interpreted in terms of quasiparticle
interference peaks corresponding to an ``octet'' of scattering wave
vectors connecting ${\bf k}$-points where the  density of states is
maximal. Until now, theoretical treatments have not been able to
reproduce the experimentally observed weights and widths of these
``octet'' peaks; in particular, the predominance of the dispersing
``${\bf q_1}$'' peak parallel to the Cu-O bond directions has remained
a mystery. In addition, such theories predict ``background''
features which are not observed experimentally.  Here, we show
that most of the discrepancies can be resolved when a realistic
model for the out-of-plane disorder in BSCCO is used. Weak
extended potential scatterers, which are assumed to represent
cation disorder, suppress large-momentum features and broaden the
low-energy ``${\bf q_7}$''-peaks, whereas scattering at order parameter
variations, possibly caused by a dopant-modulated pair interaction
around interstitial oxygens, strongly enhances the dispersing
``${\bf q_1}$''-peaks.

\end{abstract}

\pacs{74.25.Bt,74.25.Jb,74.40.+k}
\maketitle
\section{Introduction}

\begin{figure}[t]
\begin{center}
\includegraphics[width=.6\columnwidth]{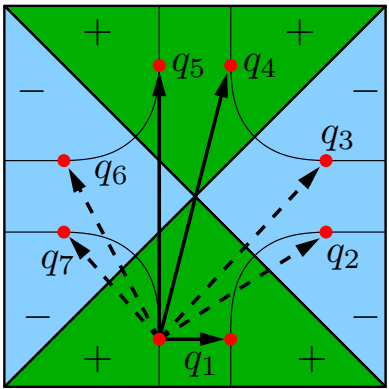}
\end{center}
\caption{Schematic depiction of the first Brillouin zone in cuprate
superconductors. Green (blue) regions represent $\Delta_{\bf k}>0$
($\Delta_{\bf k}<0$). Red dots are intersections of constant
quasiparticle energy contours with the Fermi surface. The fraction
of octet vectors, ${\bf q}_1$, ${\bf q}_4$, ${\bf q}_5$ which should in principle
be visible in the case of a ``pointlike'' (4-bond)
$\tau_1$-scatterer are denoted with solid lines, the remaining
octet vectors are indicated with dashed lines.} \label{fig:BZ}
\end{figure}

For several years, high-resolution scanning tunnelling microscopy
(STM)
experiments~\cite{yazdani,davisnative,davisZn,cren,davisinhom1,davisinhom2,Kapitulnik1,Kapitulnik2,Kapitulnik3,Hoffman1,native,McElroy,McElroy04,momono05}
have been imaging the local electronic structure of cuprate
superconductors. These experiments have discovered resonant defect
states~\cite{yazdani,davisnative,davisZn} and revealed the
existence of nano-scale
inhomogeneities~\cite{cren,davisinhom1,davisinhom2,Kapitulnik1}.
From a highly disordered spatial tunneling signal, it has become
standard to use Fourier transform scanning tunnelling spectroscopy
(FT-STS)~\cite{sprunger} techniques to extract, e.g., the
characteristic wavelengths of Friedel-type
oscillations~\cite{friedel} and, if present, the background charge
 order. In a simple metal, the local density of states
around an impurity varies as $\sim \cos 2k_F r/r^3$, so the
wavelength of the local density of states (LDOS) ``ripples" caused
by a single impurity is a measure of the Fermi wave vector of the
pure system~\cite{sprunger}. In the quasi-2D $d$-wave
superconductor, the spherical shell of maximum FT-STS intensity
with radius 2$k_F$ is replaced by a discrete set of peaks at
positions $\q_\alpha$, $\alpha=0...7$ (see Fig.~\ref{fig:BZ})
which connect an octet of points of high density of states in
momentum
space~\cite{Hoffman1,Tingfourier,WangLee,andersen,Tingfourierfewimp,Franz}.
The fact that in optimally doped BSCCO the measured dispersion of
these peaks with STM bias~\cite{Hoffman1,McElroy,McElroy04} agrees
semi-quantitatively with the predictions based on $d$-wave
BCS-theory
is considered as strong evidence for the existence of well-defined
quasiparticles in the superconducting state of optimally doped
cuprate superconductors, and for the applicability of $d$-wave
BCS-theory in this regime.  Recent attempts to relate ARPES
spectral functions to  STM Fourier transform patterns have met
with reasonable success, supporting this
conclusion~\cite{Markiewicz,Chatterjee05,McElroyARPES05}.

At lower doping, the dispersing peaks are still present, but
additional nondispersing peaks  are observed corresponding to wave
vectors along the Cu-O bonds with wavelength close to four lattice
spacings~\cite{McElroy04,momono05}. Nondispersing peaks have
been reported in overdoped samples as well, and have been discussed in terms
of stripelike charge ordering~\cite{Kapitulnik2,Kapitulnik3}.  The
nature and degree of charge ordering and the existence of the
nondispersive peaks in these materials is still controversial,
however.  In this work we neglect charge ordering and discuss
exclusively the Fourier transform LDOS response of the $d$-wave
superconducting state to disorder.

Despite the overall success of interpreting the Fourier STM
patterns in terms of quasiparticle interference, none of the existing
models has been able to correctly account for the observed weights and widths of
the dispersive $q$-space peaks.
The pointlike single-impurity
theories~\cite{Hoffman1,Tingfourier,WangLee,andersen,Tingfourierfewimp,Franz}
yield peak structures which are sharp, and linear or arclike in
structure, in contrast to the experimental features, which appear
as a series of broad, roughly round spots. Depending on the
assumed form of the impurity potential and its  strength, the
relative intensity weights of the different octet peaks change,
but there has been no choice of potential found which adequately
accounts for the experimentally observed weights. In particular,
all such calculations appear to drastically underestimate the
weight of the dispersing $\q_1$-peaks and in contrast to
experiment predict many
 ``background'' features, i.e., strong intensity in regions of
$q$-space other than the octet peaks.

Many-impurity calculations with weak scatterers displayed some of
the same features as the single-impurity calculations, and showed how
increasing disorder eventually swamps the octet peaks by raising
the noise floor, but were not notably more successful in
reproducing the experimental FT-STS patterns~\cite{capriotti}. It
was therefore suggested by Zhu {\it et al.}~\cite{ZAH04} that it
is essential to consider a more realistic disorder model for
BSCCO, including the effect of roughly 0.2\% in-plane native
defects, which are generally considered as strong pointlike
scatterers due to the resonances they generate near zero
bias~\cite{native}, as well as the effect of out-of-plane defects,
which most likely act as weak extended scatterers due to poor
screening within the CuO$_2$-planes. Using this disorder model Zhu
{\it et al.} were able to reproduce the experimental FT-STS patterns at
low biases of order 10-15meV reasonably well,  but at higher
energies agreement was rather poor. They attributed the low-energy
behavior to the in-plane native defects and interpreted the role
of the extended weak out-of-plane scatterers as broadening the
octet peaks. A broadening of small-momentum features
has also been reported by Dell'Anna {\it et al.}~\cite{DellAnna}
due to mesoscopic inhomogeneities.

In this paper we show that a realistic treatment of the out-of-plane
disorder plus the consideration of in-plane native defects
is necessary to understand the key features of the
experimental Fourier transform patterns.
There are at least two major sources of out-of-plane disorder in this
material: (i)  non-stoichiometric oxygen dopant atoms and (ii)
random substitution of Bi at the Sr-sites~\cite{Eisaki}. Very
recently, direct information on both types of impurities has been
obtained by STM.  Kinoda {\it et al.}~\cite{Kinoda05} succeeded in
imaging the excess Bi$^{3+}$  at a bias of 1.7~eV, finding an
areal concentration of about 3\%, consistent with earlier estimates
from bulk measurements~\cite{Eisaki}.  In addition, McElroy {\it et
al.}~\cite{DavisScience05} imaged a set of atomic scale defects at
-0.96 eV, whose concentration scales with
doping. These defects, which  were  ultimately identified as the interstitial oxygen
dopant atoms, were shown to correlate strongly and
positively with the local gap size in the superconducting state.

We have recently suggested that the above mentioned experimental
observations can be understood based on the simple assumption that
the main effect of the interstitial dopants
(i.e. disorder (i)) is to increase the
pair interaction locally~\cite{NAMH05}.
This assumption yields
gap modulations which correlate positively with the dopant
positions, gives rise to particle-hole symmetric spectra and
suppresses high-energy coherence peaks as observed in experiment.
The underlying assumption is that each  interstitial oxygen  distorts the
``cage" of atoms around it, thereby modulating the pair
interaction locally.
Alternatively, these order parameter modulations could also arise from
sources other than microscopic pairing disorder, e.g. via coupling of
ordinary electrostatic impurity potentials to phases which compete
with $d$-wave superconductivity, see e.g. Ref.~\cite{chakravarty}.
Nevertheless, the phenomenological assumption of microscopic
pairing disorder has already proven successful with respect to
many experimental observations in BSCCO. Scattering by order
parameter variations has been reported to improve dramatically the
fit to the inelastic tunneling spectroscopy
experiments~\cite{BalatskyIETS}. Fang {\it et al.}~\cite{Fang}
reported very sharp coherence peaks in small gap regions
consistent with an Andreev bound state in a region of suppressed
order parameter, as discussed in Ref.~\cite{NAMH05}. Very
recently, J.-X. Zhu explored the effect of dopant modulated Cu-Cu
hopping within a Bogoliubiov-de Gennes framework~\cite{JXZhu05}.
Finally, order parameter modulations with a phase twist have been
found to yield low-energy impurity bound state energies and a
spatial distribution of the LDOS  similar to those observed near
Zn-impurities in BSCCO-2212~\cite{phaseimp}. Thus it is intriguing
to explore the consequences of this unusual quasiparticle
scattering from the order parameter modulations further, in
particular its effect on the FT-STS patterns.

Unlike the oxygen dopants, the Bi$^{3+}$ defects, i.e.
disorder component (ii), most likely act
as conventional weak and extended potential scatterers due
to the different charge of the Sr- and Bi-ions and the smaller
deformation of the lattice itself.
The existence of a large concentration of
extended weak potential scatterers is also supported by an
analysis of transport properties in BSCCO~\cite{microwave}.

The final aspect of the disorder we discuss  is the small
concentration of in-plane native defects, about 0.2\% per Cu,
imaged by STM~\cite{native}. It is believed that these are Cu
vacancies in the CuO$_2$ plane, although  no direct proof exists.
Individually, these defects provide a resonant signal centered
within 1 meV of the Fermi level similar but not identical to Zn.
The resonance modulates the LDOS primarily below 15 meV or less,
and then furthermore at energies not far from the gap edge where
the resonant spectral weight originates.  We model these defects
as strong pointlike unitary potential scatterers, as in Zhu {\it et
al.}~\cite{ZAH04}, noting that because the concentration and resonance
energies of these defects are known from experiment, they do not
introduce new model parameters.

Thus, we believe that a realistic model of the disorder in BSCCO
should contain a net out-of-plane disorder consisting of
extended weak potential scatterers (approx. 3\%, i.e., the
concentration of Bi-defects) and extended off-diagonal
scatterers (approx. 7.5\%, i.e., half the doping, because
each oxygen dopant most likely contributes two holes) plus a
concentration of 0.2\% pointlike and strong in-plane defects. It is
the purpose of this paper to investigate the effect of these three
disorder components on the FT-STS patterns.

\section{Model}
As a model for the homogeneous system we use the usual BCS
Hamiltonian given by
\begin{equation}
{\mathcal{H}}_0=\sum_{{\mathbf{k}},\sigma}
\varepsilon_{\mathbf{k}} \hat{c}^\dagger_{{\mathbf{k}}\sigma}
\hat{c}_{{\mathbf{k}}\sigma} + \sum_{{\mathbf{k}}} \left(
\Delta_{\mathbf{k}} \hat{c}^\dagger_{{\mathbf{k}}\uparrow}
\hat{c}^\dagger_{-{\mathbf{k}}\downarrow} + {\rm H.c.} \right),
\end{equation}
where $\varepsilon_{\k}$ is the quasiparticle dispersion proposed to
fit the ARPES data~\cite{norman}: $\varepsilon_{\k}=\sum_{n=0}^5
t_n\chi_n(k)$, where $t_{0-5}=0.1305$, $-0.5951$, $0.1636$,
$-0.0519$, $-0.1117$, $0.0510$ (eV), and $\chi_{0-5}(k)=1$,
$(\cos k_x+\cos k_y)/2$, $\cos k_x\cos k_y$, $(\cos 2k_x+\cos
2k_y)/2$, $(\cos 2k_x\cos k_y+\cos 2k_y\cos k_x)/2$, and $\cos
2k_x\cos 2k_y$. For $d$-wave pairing we have,
$\Delta_{\mathbf{k}}=2\Delta_0 (\cos k_x - \cos k_y)$
with $\Delta_0=12{\rm meV}$.

In terms of the Nambu spinor
$\hat{\psi}^\dagger_{{\mathbf{k}}}=(\hat{c}^\dagger_{{\mathbf{k}}\uparrow},
\hat{c}_{-{\mathbf{k}}\downarrow})$,
the corresponding bulk Green's function in Matsubara
representation is given by
\begin{equation}\label{nambugreensfunctions}
{\mathcal{G}}^{0}({\mathbf{k}},i\omega_n)=\frac{i\omega_n\tau_0+\xi_{\mathbf{k}}\tau_3
+\Delta_{\mathbf{k}}\tau_1}{(i\omega_n)^2-E_{\mathbf{k}}^2},
\end{equation}
where $E_{\mathbf{k}}^2=\xi_{\mathbf{k}}^2+\Delta_{\mathbf{k}}^2$,
and $\tau_0$ is the $2\times 2$ identity matrix, whereas
($\tau_1,\tau_2,\tau_3$) denote the three Pauli matrices. In
real-space, the perturbation due to the conventional point-like
potential scatterers in the diagonal $\tau_3$ channel and
modulated pairing amplitude in the off-diagonal $\tau_1$ channel
can be written as
\begin{equation}
{\mathcal{H}}^{\prime}({\mathbf{r}},{\mathbf{r}}')=\hat{\psi}^\dagger_{{\mathbf{r}}}
\left[ V({\mathbf{r-r'}})\delta({\mathbf{r-r'}}) \tau_3 + \delta
\Delta({\mathbf{r}},{\mathbf{r}}') \tau_1 \right]
\hat{\psi}_{{\mathbf{r}}'}.\label{Hprime}
\end{equation}
Note that
the conventional impurity scattering occurs with Pauli matrix
$\tau_3$ in (\ref{Hprime}), whereas the order parameter
modulation, or Andreev scattering term,  occurs with $\tau_1$.  We
will henceforth adopt the convenient shorthand of referring to the
two types of processes as $\tau_3$ or $\tau_1$ disorder,
respectively.

We use a Yukawa form to model the extended conventional potential
of an out-of-plane defect:
\begin{equation}
V({\bf r})=\frac{V_0 r_z}{e^{-r_z/\lambda}} \frac{e^{-d/\lambda}}{d}
\quad {\rm with} \quad d=\sqrt{r^2+r_z^2} \,,
\end{equation}
where ${\bf r}$ is an in-plane vector, $r_z$ is the distance of the
defect from the CuO$_2$-plane and the potential is normalized to
$V_0$ in the CuO$_2$-plane directly below the defect. An extended
$\tau_1$-scatterer is modeled by using the same functional form
in the $\tau_1$ channel as
\begin{equation}
\delta \Delta ({\bf r}, {\bf r}')= \frac{\delta \Delta \, r_z}{e^{-r_z/\lambda}}
\frac{e^{-d/\lambda}}{d}\,\, {\rm with} \,\, d=\sqrt{(\frac{1}{2}({\bf r}+{\bf r}'))^2+r_z^2}\,.
\end{equation}
We note that there is no microscopic justification for this form,
nor any direct connection with a physical screened Coulomb
potential; this is merely a convenient way to introduce a length
scale to the order parameter modulations induced by the dopant
atoms.

In order to determine the resulting LDOS as a function of energy
and lattice sites, one needs to obtain the full Green's function
${\mathcal{G}}({\mathbf{r}},{\mathbf{r}}',i\omega_n)$ given by the
Dyson equation
\begin{equation}\label{dyson}
{\mathcal{G}}({\mathbf{r}},{\mathbf{r}}')={\mathcal{G}}^{0}({\mathbf{r}}-{\mathbf{r}}')
+\sum_{{\bf r}'',{\bf r}'''} {\mathcal{G}}^0({\mathbf{r}}-{\mathbf{r}}'')
H^{\prime}({\mathbf{r}}'',{\mathbf{r}}''')
{\mathcal{G}}({\mathbf{r}}''',{\mathbf{r}}').
\end{equation}
Thus, by calculating
\begin{equation}\label{G0realspace}
{\mathcal{G}}^{0}({\mathbf{r}},i\omega_n)= \sum_{\mathbf{k}}
\frac{(i\omega_n\tau_0+\xi_{\mathbf{k}}\tau_3+\Delta_{\mathbf{k}}\tau_1)}{(i\omega_n)^2-E_{\mathbf{k}}^2}
\exp(i {\mathbf{k}} \cdot {\mathbf{r}}),
\end{equation}
the remaining problem is that of a simple matrix inversion, and
the LDOS $\rho({\bf r},\omega)$ can be extracted from the imaginary
part of the full Green's function, $\rho({\bf r}, \omega) =
-\frac{1}{\pi} \mbox{Im} {\mathcal G}_{11} ({\bf r}, {\bf r}, \omega)$.

\begin{figure*}
\begin{center}
\includegraphics[width=.99\textwidth,bb=50 210 560 750]{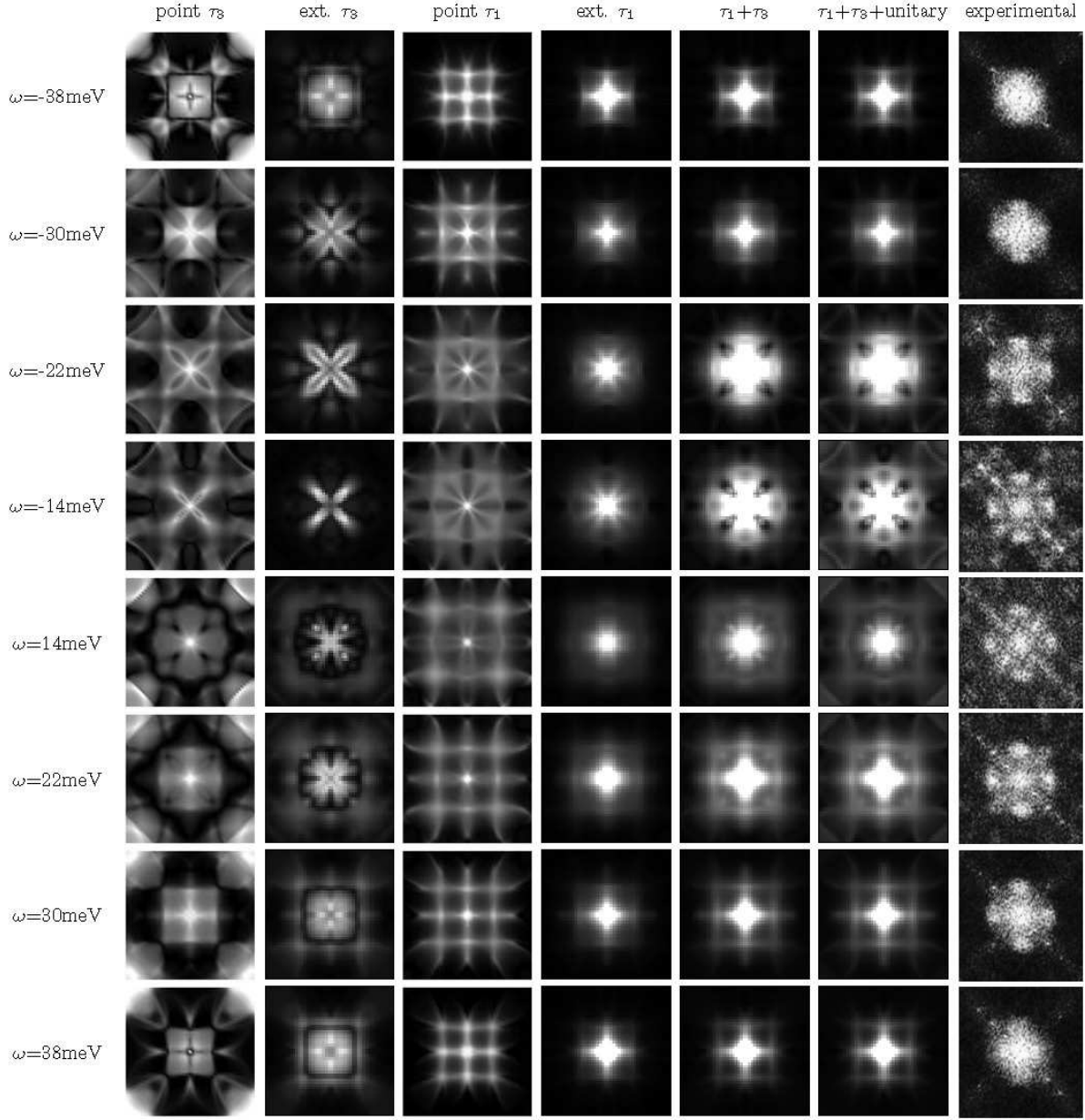}
\end{center}
\caption{Effect of out-of-plane scattering potentials of different
type on the power spectrum.  Shown is the power spectrum for: 1st
column: pointlike $\tau_3$-scatterer with strength $V$=120meV, 2nd
column: extended $\tau_3$-scatterer with $V_0$=40meV,
   $\lambda=r_z=1.5a$,
3rd column: pointlike $\tau_1$-scatterer with $\delta
\Delta=\Delta_0$
   on the four central bonds,
4th column: slightly extended $\tau_1$-scatterer with
$\delta \Delta=\Delta_0$,
   $\lambda=r_z=1.2a$,
5th column: combination of 7.5\% extended $\tau_1$-scatterers
   ($\delta \Delta=\Delta_0$, $\lambda=r_z=1.2a$) and 3\% extended
   $\tau_3$-scatterers ($V_0$=40meV, $\lambda=r_z=1.5a$),
6th column: same as 5th column plus 0.2\% unitary ($V$=3.9meV)
  pointlike scatterers,
7th column: experimental power spectrum~\cite{McElroy,McElroy04}.}
\label{fig:FTdos}
\end{figure*}

\section{Results}

 We now examine the effect of a single scatterer
on the FT-STS patterns.  The philosophy of calculations of this
type~\cite{WangLee,Tingfourier} is simply that the ``Friedel
oscillations" in the $d$-wave quasiparticle sea around a single
impurity will contain many of the wavelengths present in the fully
disordered system. The FTDOS due to the impurity is then given by
\begin{equation}
   \rho(\q,\omega) = \sum_{{\bf r}\in L\times L} e^{-i\q\cdot {\bf r}} \rho({\bf r}, \omega)
 \end{equation}
where $L\times L$ is a square set of $L^2$ positions at which
measurements are made, and $\q = 2\pi(m,n)/L$ are vectors in the
associated reciprocal lattice. The power spectrum~\cite{footnote}
is simply the 2D image obtained from the amplitude
$|\rho(\q,\omega)|$. In the case that several disorder components
are present we calculate the power spectrum as
$|\rho(\q,\omega)|=|\sum_i n_i \rho_i (\q,\omega)|$, where
$\rho_i$ and $n_i$ denote the FTDOS and concentration of each
disorder component.

Fig.~\ref{fig:FTdos} compares the resulting power spectra for
different impurity models computed for $L=53$. In the 1st column,
the power spectrum for a weak pointlike $\tau_3$-impurity,
discussed previously in Refs.~\cite{WangLee,ZAH04}, is reproduced.
In contrast to the experimental power spectrum, the calculation
for a pointlike $\tau_3$-impurity predicts strong weight at large
wavevectors especially near $(\pi,\pi)$, which is observed
experimentally only at very low frequencies. In the case of an
extended $\tau_3$-scatterer (2nd column in Fig.~\ref{fig:FTdos}),
on the other hand, the high-momentum features are suppressed due
to the rapid fall-off  of the impurity potential $V({\bf q})$ in
momentum space. This is, however, not the only difference between
pointlike and extended $\tau_3$-scatterers. The FT-patterns also
change for small momenta; most noticeably, the $\q_7$-peaks are
enhanced due to small-angle scattering of quasiparticles by the
smooth potential. Whereas the $\q_7$-peaks appear only as tiny
spots in the case of a pointlike impurity~\cite{ZAH04}, too small
to be resolved in Fig.~\ref{fig:FTdos}, they become broad spots in
the case of an extended impurity and are clearly visible at the
energies $|\omega|=14$meV, 22meV in Fig.~\ref{fig:FTdos}. This
supports the proposal by Zhu {\it et al.}~\cite{ZAH04} that the
enhancement of forward scattering due to spatially extended
impurities broadens the octet peaks.

In addition to the existence of weak extended $\tau_3$-scatterers,
which could arise e.g. due to random substitutions of Sr by Bi,
the pair interaction is most likely modulated due to the presence
of non-stoichiometric oxygen dopant atoms~\cite{NAMH05}. This
results in inhomogeneities in the magnitude of the superconducting
order parameter and enhanced Andreev scattering which can be
modeled as $\tau_1$-scattering within a single-impurity approach.
The 3rd column in
Fig.~\ref{fig:FTdos} shows the power spectrum for a ``pointlike''
$\tau_1$-scatterer, i.e., a modulation of the order parameter on
the four bonds surrounding the impurity site as $\delta \Delta
(0,\pm \hat x)=\delta \Delta (\pm \hat x,0)= -\delta \Delta (0,\pm
\hat y)=-\delta \Delta (\pm \hat y, 0)=\delta \Delta$. Obviously,
the resulting FT-pattern looks very different from a pointlike
$\tau_3$-scatterer. The background features, i.e., the finite
FTDOS at wavevectors that do not correspond to one of the octet
peaks, are much less pronounced. At high energies
($|\omega|=30$meV, 38meV), the power spectrum is clearly dominated
by $\q_1$-peaks, whereas the $\q_7$-peaks are completely
absent even at the lowest energies. This behavior can be
understood in terms of the momentum dependence of the scattering
matrix element for a ``pointlike'' $\tau_1$-scatterer. Fourier
transformation of the ``pointlike'' order parameter modulation
yields
\begin{equation}
\delta\Delta_{\k\k'} = \delta\Delta (\Delta_\k+\Delta_{\k'})/\Delta_0 \,,
\end{equation}
i.e., $\delta\Delta_{\k\k'}$ vanishes for wavevectors $\k$ and
$\k'$ which connect points of the Brillouin zone where $\Delta_\k$
and $\Delta_{\k'}$ have opposite signs but equal magnitudes. Among
the octet vectors, only $\q_1$, $\q_4$ and $\q_5$ are allowed
scattering processes contributing to the  FTDOS for a
``pointlike'' $\tau_1$-scatterer, whereas $\q_2$, $\q_3$, $\q_6$
and $\q_7$ are suppressed because the matrix element
$\delta\Delta_{\bf k,k'}$ vanishes for these $\q$, see
Fig.~\ref{fig:BZ}. The 4th column of Fig.~\ref{fig:FTdos} shows
the power spectrum for the case of a slightly more extended
$\tau_1$-scatterer. The primary effect of extending the spatial
range of the order parameter modulation is analogous to the $\tau_3$-case, i.e.,
the intensity at large momenta and background features are
suppressed resulting in even more pronounced $\q_1$-peaks.

The 5th column of Fig.~\ref{fig:FTdos} presents the power spectrum
of an extended impurity with a combined $\tau_1$- plus
$\tau_3$-potential, where each single component corresponds to the
potential used for the extended $\tau_3$/$\tau_1$-case in the 2nd
/4th column. Since we assume that the extended $\tau_3$-scatterers
result from the 3\% Sr/Bi-disorder and the extended
$\tau_1$-scatterers from the 7.5\% dopant oxygens the FTDOS of
each component has been weighted with a prefactor corresponding to
the ratio of these two concentrations. Alternatively, a scatterer
with combined $\tau_1$- and $\tau_3$-character could also arise
due to the fact that each oxygen dopant, besides modulating the
pair potential, most likely also possesses a small but finite
conventional potential component.
 The extended
$\tau_1+\tau_3$-case obviously combines the main characteristics
of the extended $\tau_1$ and the extended $\tau_3$-case. At
$|\omega|=14$meV the power spectrum is dominated by $\q_7$-peaks, at
$|\omega|=22$meV both the $\q_7$- and the
$\q_1$-peaks are present, whereas the highest frequencies $|\omega|=30$meV and
$|\omega|=38$meV are clearly dominated by $\q_1$-peaks.

\begin{figure}
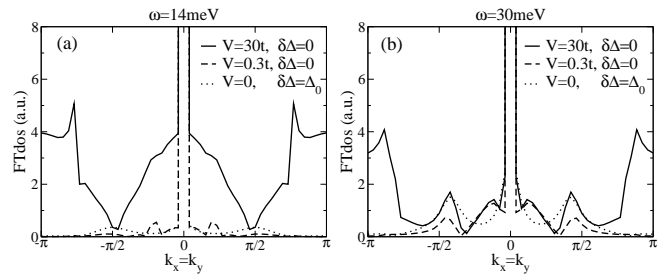

\begin{center}
\begin{minipage}{.49\columnwidth}
\includegraphics[width=1.0\columnwidth,clip=]{LinescanDiagw014.eps}
\end{minipage}
\begin{minipage}{.49\columnwidth}
\includegraphics[width=1.0\columnwidth,clip=]{LinescanDiagw030.eps}
\end{minipage}
\end{center}
\caption{Cut along the diagonal of the Brillouin zone ($k_x$=$k_y$)
through the power spectrum of a strong
pointlike $\tau_3$-scatterer with $V$=3.9eV (solid line), a weak
extended $\tau_3$-scatterer with $V_0$=40meV, $\lambda=r_z$=1.5a (dashed
line) and a ``pointlike'' $\tau_1$-scatterer with $\delta \Delta=\Delta_0$
(dotted line) for an energy of (a) $\omega$=14meV and (b) $\omega$=30meV.}
\label{fig:LinescanDiag}
\end{figure}

In addition to out-of-plane impurities, a realistic description of
the disorder in BSCCO has to take into account the in-plane native
defects with concentration estimated as 0.2\% from STM. These
defects are usually modeled as pointlike unitary impurities
located within the CuO$_2$-planes. The 6th column of
Fig.~\ref{fig:FTdos}
 displays the power spectrum resulting from the combined effect
 of the extended $\tau_1$- and $\tau_3$-scatterers, as
displayed in the 5th column of Fig.~\ref{fig:FTdos}, plus 0.2\%
pointlike unitary impurities (here $V$=3.9eV). The
existence of native defects within the CuO$_2$-planes of BSCCO is
only of minor importance for the high-frequency power spectrum,
because their concentration is much smaller than that of the
out-of-plane defects. At smaller frequencies, however, the strong
pointlike impurities produce additional large-momentum features,
especially close to $(\pi,\pi)$. This is precisely what is
observed in  experiment:  large-momentum octet vectors are only
observed at low frequencies.
The reason why the large-momentum features
appear only at low energies can be understood from
Fig.~\ref{fig:LinescanDiag}, where the power spectrum along the
Brillouin zone diagonal is compared for a single unitary pointlike
scatterer with the power spectrum resulting from a single weak
extended scatterer.
Note, that the intensity of  the large-momentum peak, which results
from the pointlike unitary scatterers, is almost frequency
independent (see Fig.~\ref{fig:LinescanDiag}). However, its weight
relative to the small-momentum features of FTSTS signal  decreases
for large bias values. This occurs mainly because the intensity of
the low momentum features, contributed by the weak scatterers,
grows rapidly with frequency.

Overall, the power spectrum calculated within our realistic
disorder model (6th column in Fig.~\ref{fig:FTdos}) reproduces the
experimental one (7th column in Fig.~\ref{fig:FTdos}) very well.
This allows us to interpret some of the key features of the
experimental power spectrum in the following way: (i) the dominant
$\q_1$-peaks arise mainly from scattering by the  order parameter
 modulations, i.e., by $\tau_1$-scattering,
(ii) the broadness of the $\q_7$-peaks, which are quite pronounced
at small energies, is caused by poorly screened out-of-plane
potential impurities, which act as extended $\tau_3$-scatterers,
(iii) the large momentum features at small frequencies originate
from the in-plane native defects, which are modeled as pointlike
unitary scatterers.

\section{ Conclusions}

We have suggested that a realistic model for the out-of-plane
disorder in BSCCO-2212 should include: (i) order parameter
inhomogeneities, which we have previously attributed to the fact
that the oxygen dopant atoms modulate the pair interaction
locally~\cite{NAMH05} and which result in enhanced Andreev or
$\tau_1$-scattering, (ii) extended weak potential  or
$\tau_3$-scatterers, which could originate e.g. from the random
substitution of Sr by Bi and which are spatially extended due to
poor screening within the CuO$_2$-planes. In combination with the
0.2\% in-plane native defects observed by STM, which we model as
pointlike unitary potential scatterers, our model for the
out-of-plane disorder is able to reproduce the essential
characteristics of the experimental Fourier transform patterns in
BSCCO.
The fact that the out-of-plane scatterers are spatially extended
explains the absence of large-momentum octet peaks and the
suppression of ``background'' features at higher energies. The
large-momentum features at small energies are caused by the 0.2\%
pointlike strong in-plane scatterers. The $\q_7$-peaks are
broadened considerably due to enhanced forward scattering caused
by the spatially extended potential impurities. The
$\tau_1$-scattering, which arises from the order parameter
inhomogeneities observed in BSCCO, strongly enhances the
dispersive $\q_1$-peaks especially at higher energies. The reason
why $\tau_1$-scattering mostly contributes to $\q_1$-peaks is a
simple consequence of the $d$-wave symmetry of the corresponding
scattering matrix elements. \vskip .2cm

The agreement of the qualitative features of the power spectrum
predicted by the model introduced in Ref.~\cite{NAMH05} for the
out-of-plane O dopants with experiment is further evidence in
support of the contention put forward there that the O defects in
the BSCCO-2212 system are actually modulating the pair interaction
$g$ locally.  Of course comparisons of this type cannot directly
determine the microscopic origin of the modulation of $g$, but
since the experimental data with which we compare are from
slightly overdoped samples, it seems unlikely that significant
inhomogeneity is being driven by  competition with a second order
parameter.  This therefore suggests that an atomic-scale
modulation of electronic hopping parameters or electron-phonon coupling
constants is indeed taking place in these materials near the dopant
atoms.\cite{NAMH05}.

\vskip .2cm

{\it Acknowledgements.} The authors acknowledge valuable
conversations with W.A. Atkinson,  A.V. Balatsky, J.C. Davis,  J. Lee and
K. McElroy.  We would especially like to thank  J.C. Davis, J. Lee and K.
McElroy for sharing with us their data.
Partial support for this research (BMA, TSN and PJH) was provided
by ONR N00014-04-0060. TSN was supported by a Feodor-Lynen Fellowship
from the A. v. Humboldt Foundation, and AM by the Institute for Fundamental
Theory of the University of Florida.


\begin{thebibliography}{scalapino}
\bibitem{footnote} Note we use the term ``power spectrum" for the
square root of the conventional definition.
\end{thebibliography}
\end{document}